\documentstyle[aps,pra,epsf]{revtex}

\begin{document}
\draft
\title{\bf Effect of an inhomogeneous external magnetic field on a quantum
dot quantum computer}
\author{Rogerio de Sousa, Xuedong Hu, and S. Das Sarma}
\address{Department of Physics, University of Maryland, College Park, MD 
20742-4111}
\date{\today}
\maketitle

\begin{abstract}
We calculate the effect of an inhomogeneous magnetic field, which is
invariably present in an experimental environment, on the exchange energy
of a double quantum dot artificial molecule, projected to be used as a
2-qubit quantum gate in the proposed quantum dot quantum computer. We use
two different theoretical methods to calculate the Hilbert space structure
in the presence of the inhomogeneous field: the Heitler-London method
which is carried out analytically and the molecular orbital method which
is done computationally. Within these approximations we show that the
exchange energy J changes slowly when the coupled dots are subject to a
magnetic field with a wide range of inhomogeneity, suggesting swap
operations can be performed in such an environment as long as quantum
error correction is applied to account for the Zeeman term. We also point
out the quantum interference nature of this slow variation in exchange.
\end{abstract}

\section{Introduction}

It has long been proposed that quantum mechanical computers might provide
much more computing power than machines based on classical
physics\cite{feynman}. However, it was the discovery of fast quantum
algorithms for tasks such as prime factorization and searching disordered
databases\cite{grover} together with effective error correction
schemes\cite{shor} that motivated a frenetic search for a suitable
physical system on which reliable quantum hardware could be built.

Among the many proposed quantum hardware schemes the ion trap quantum
computer, based on electronic states of laser cooled trapped ions, have
not only demonstrated full control of elementary one and two-qubit
(quantum bit, the basic unit of quantum information) logic gates, but also
recently produced 4-qubit entanglement\cite{wineland}. Although the ion
trap represents the current state of the art in quantum computing
architecture, its scalability prospect is unclear, since moving a large
number of ions between memory and computing sectors of the ion traps
without destroying quantum coherence is a difficult problem. Another
proposed scheme, the liquid state Nuclear Magnetic Resonance (NMR) quantum
computer, based on nuclear spin states of molecules in solutions, recently
demonstrated a 5-qubit order finding algorithm\cite{vandersypen}. However,
the NMR quantum computer also has a scalability problem, since the signal
to noise ratio decreases exponentially as the size of the molecules
increases. Furthermore, there has been a theoretical proof that no
entanglement is present in the current ensemble-averaged NMR
experiments\cite{braunstein}, although its implication for NMR quantum
computation has not yet been fully explored.

With the scalability problem in mind it has been suggested that advanced
semiconductor technology might provide the basis for building a large
(i.e. many qubits) quantum computer (QC), just as it has been proved to be
effective in building today's classical digital computers. Two of the main
proposals for such semiconductor QCs are based on qubits using the spin of
a phosphorous donor nucleus in bulk silicon\cite{kane} and the spin of a
single electron trapped in a GaAs quantum dot\cite{loss}. Even though our
work here refers specifically to the latter, our results are valid for
both solid state QC proposals since the two are similar in the sense that
they use the same universal quantum gate to perform quantum computation:
The exchange or swap gate which acts on two qubits coupled by an effective
Heisenberg spin interaction and one-qubit gates that perform rotations on
two orthogonal spin axes by applying external magnetic fields.
Experimentally demonstrating the operation of any of these quantum gates
is a formidable task and has not been carried out yet, and all
semiconductor based QCs are at best promising proposals at this stage.
Nevertheless, the hope is that after the first step of successfully
demonstrating a single qubit in a semiconductor-based QC, scaling up from
one qubit to many would be feasible based on the established
infrastructure of semiconductor microelectronics industry. Moreover,
precisely controlling the quantum dynamics of a few spins in a
semiconductor nanostructure represents a major experimental challenge, and
therefore any success will lead to new insight into mesoscopic spin
properties of nanostructures, which by itself is a subject of intrinsic
fundamental interest.

Quantum computation can be considered as a complicated unitary evolution
of a N-qubit system, governed by a N-qubit Hamiltonian, where each qubit
is a quantum mechanical two-level system. It has been proved that this
usually extremely complicated process can be factored into a number (that
is a polynomial in N) of simple one and two-qubit processes with
sufficiently high accuracy, in analogy with a classical network of logic
gates\cite{div1}. To improve efficiency multiple simple operations might
be lumped together, as was proposed in various parallel pulse
schemes\cite{guido2}. In these operations for a quantum dot quantum
computer (QDQC) an inhomogeneous magnetic field inevitably appears, as it
is a powerful technique to rotate spins while controlling spin exchange.
The inhomogeneity in the field here arises necessarily from the fact that
the parallel scheme employs local magnetic field pulses on individual dots
to achieve different spin rotations, and a magnetic field localized on one
dot is essentially an inhomogeneous field for a different dot.
Furthermore, inhomogeneous magnetic fields can be produced by unwanted
magnetic impurities or external currents in the sample, which will be
invariably present in any real QDQC architecture. Since in the proposed
parallel schemes the exchange coupling J (which controls the swap gate) is
treated as independent of the magnetic field inhomogeneity\cite{guido2},
it is important to explore whether this dependence is truly weak enough,
or if it is necessary to adjust the inter-dot barrier continuously in
order to maintain the desired J in the presence of a varying inhomogeneous
field (which will be a formidable, if not impossible, task).

We showed in a previous paper\cite{zeeman} that an inhomogeneous magnetic
field leads to errors in the swap operation of two qubits because the
Zeeman term in the effective Hamiltonian is not proportional to the total
spin. For example, if we try to swap the state $\left| \uparrow \downarrow
\right\rangle$ we would obtain $\left| \downarrow \uparrow \right\rangle$
plus an entangled component proportional to the magnetic field
inhomogeneity. In other words, once a product state is entered, the output
will always be entangled, although for practical purposes this error can
be kept within the currently estimated error correction bounds ($\simeq
10^{-4}$) for usual field inhomogeneities. This result is based on the
assumption that one knows J exactly in order to achieve the best control,
otherwise there can be additional errors in swap because of the
uncertainty in J.

Motivated by the above concerns and problems arising from inhomogeneous
magnetic fields, we calculate in this paper the exchange coupling J of a
two-electron double dot as a function of the inhomogeneity in the external
magnetic field. We employ a single envelope function approach and
calculate the orbital energies of the two electrons, thus relating the
general two-electron problem to a special two-spin Hamiltonian, and
analyze the effect of field inhomogeneities on the orbital degrees of
freedom of the double dot structure. The current work is thus
complementary to our recent publication\cite{zeeman} dealing with the
field inhomogeneity induced errors in the swap operation.

Quite recently it has been demonstrated that QDQC can be done without any
applied magnetic field as long as a price is paid in increasing the number
of exchange operations: Two ($S=\frac{1}{2},S_{z}=+\frac{1}{2}$) states of
three spins are used as a qubit, and the nearest-neighbor exchange
interaction can provide all the required one-qubit and two-qubit
operations of a quantum computer\cite{bacon}. In this scheme applications
of external magnetic fields for single qubit operations is not required,
as exchange gates provide both 1- and 2-qubit operations. This proposal
removes the problem of having strong applied inhomogeneous fields.
However, stray fields due to impurities and external currents can still be
present, and must be taken into account using appropriate error
corrections. In this paper we focus on clarifying the issues that arise in
the simpler sequential scheme that may contain an applied inhomogeneous
magnetic field as well as pointing out potential errors in all exchange QC
schemes. Our results are also directly relevant to the proposed parallel
pulse schemes as discussed above.

\section{Model}

We consider two electrons in a horizontally coupled GaAs double quantum dot
structure\cite{guido1,xuedong} (Fig. 1). Since these electrons are at the
bottom of
the conduction
band the single envelope function approach provides a sufficiently complete
physical picture of their energy states. Suppose there is an
inhomogeneous magnetic field applied so that the electron in the left well
is subject to a field which is on average $2B_{dif}$ lower than the one in
the right. For simplicity we assume a linearly changing field profile 
\begin{equation}
{\mathbf B}(x,z)=(\overline{B}+B_{dif}
\frac{x}{a_{m}}){\mathbf \widehat{z}}+B_{dif} 
\frac{z}{a_{m}}{\mathbf \widehat{x}}, \label{bfield}
\end{equation}
where $\overline{B}=\frac{1}{2}(B_{R}+B_{L})$,
$B_{dif}=\frac{1}{2}(B_{R}-B_{L})$, and $2a_{m}$ 
is the separation between the dots. The complete Hamiltonian is given by 
\begin{equation}
H=H_{orb}+H_{Z},  \label{htotal}
\end{equation}
\begin{equation}
H_{orb}=h_{1}+h_{2}+C  \label{horb}
\end{equation}
\begin{equation}
h_{i}=\frac{1}{2m}\left( {\mathbf p}_{i}+ 
\frac{e}{c} {\mathbf A}\right) ^{2}+V(
{\mathbf x}_{i}),
\end{equation}
\begin{equation}
C=\frac{e^{2}}{\varepsilon }\frac{1}{\left| {\mathbf x}_{1}-{\mathbf x}
_{2}\right| },
\end{equation}
\begin{equation}
H_{Z}=g^{*}\mu _{B}\sum\limits_{i=1}^{2}{\mathbf B}({\mathbf x}_{i})\cdot 
{\mathbf S}_{i} .  \label{hzeeman}
\end{equation}
Here ${\mathbf x}_{i}$ is the position of the $i$th electron in the xy
plane, $m=0.067m_{e}$ is the electron effective mass in GaAs, $\varepsilon
=13.1$ is the
dielectric constant, $g^{*}=-0.44$ is the GaAs gyromagnetic ratio, and $
\mathbf{A}$ is the vector potential for the inhomogeneous field: 
\begin{equation}
{\mathbf A}=-\left( \frac{\overline{B}}{2}y\right){\mathbf \widehat{x}}+\left[
\frac{
\overline{B}}{2}x+\frac{B_{dif}}{2a_{m}}(x^{2}-z^{2})\right] \;{\mathbf \widehat{y}}.
\label{vecpot}
\end{equation}
Note that since the quantum dots are fabricated using a high mobility 2D
electron gas (taken to be along the xy plane), the dot dynamics is assumed
to be constrained to the $z=0$
plane because the z-width of the 2D electron gas is typically much
smaller than the QD confinement in the x-y plane. The z-dependent
components in $\mathbf{A}$ and $\mathbf{B}$ do not have any effect other
than\ to guarantee the $\mathbf{\nabla }\times \mathbf{B}=0$ condition. Hence the effect of 
$B_{x}=B_{dif} z/a_{m}$
will be neglected in our calculation (the 2D electron gas has a typical
thickness of the order of a few nm and neglecting $B_{x}$ is an
excellent approximation for realistic QD systems). The confining potential
$V$ in the x-y plane is modeled in our calculations by a linear
combination of gaussians characterized by the parameters $V_{0}$, $a$, 
$V_{b}$, $l_{x}$, $l_{y}$, $l_{bx}$ and $l_{by}$ which can be adjusted to
control the size of the dots and switch on and off the exchange coupling: 
\begin{equation}
V=-V_{0}\left[ \exp \left( -\frac{(x-a)^{2}}{l_{x}^{2}}\right) +\exp
\left( -\frac{(x+a)^{2}}{l_{x}^{2}}\right) \right] \exp \left( -\frac{y^{2}}{
l_{y}^{2}}\right)+V_{b}\exp \left( -\frac{x^{2}}{l_{bx}^{2}}\right) 
\exp \left( -\frac{y^{2}}{l_{by}^{2}}\right) . \label{confpot}
\end{equation}
In our calculation we choose $V_{0}=50meV$ as the potential well depth and $
V_{b}=30meV\,$as the central barrier height. In addition, $a=15nm$, $l_{x}=$ 
$l_{y}=30nm$, $l_{bx}=15nm$, and $l_{by}=80nm$, which leads to an effective
barrier height (difference in energy between the top of the central barrier
and the bottom of one of the wells) of 9.65meV, and a slightly shifted
potential minimum $a_{m}=16.41nm$. We note that our choice of
Eq. (\ref{confpot}) as the QD confinement potential model is actually
quite a general approximation, and allows considerable flexibility in our
QDQC Hilbert space structure calculations. 

An important feature of the Zeeman Hamiltonian (\ref{hzeeman}) is that it
couples coordinate $\mathbf{x}_{i}$ with spin $\mathbf{S}_{i}$ in an
inhomogeneous field. This interaction introduces a coupling
between the
singlet and the $S_{z}=0$ triplet state, and can be well described (in the
Heitler-London approximation) by an effective spin Hamiltonian\cite{zeeman} 
\begin{equation}
H_{Z}\cong \gamma _{R}S_{1z}+\gamma _{L}S_{2z},\;\gamma _{R}=g^{*}\mu
_{B}B_{R},\;\gamma _{L}=g^{*}\mu _{B}B_{L}.  \label{hzeff}
\end{equation}
But note that Eq. (\ref{hzeff}) is not invariant under particle
permutation $P_{12}$, which is a natural consequence of the fact that it has a
non-vanishing matrix element between singlet and triplet states (which are
respectively odd and even under a $P_{12}$ operation acting only in the
two-spin space). This lifting of the permutation symmetry is a direct
consequence of the inhomogeneous external magnetic field in this problem.

Our main objective is to study the dependence of the exchange coupling J
on the inhomogeneity of the magnetic field. The parameter J describes the
splitting between the lowest energy singlet and triplet states of the
Hamiltonian (\ref{htotal}), which at low energy reduces to the effective
two-spin Heisenberg-type Hamiltonian that enables the operation of the
quantum gates, 
\begin{equation} 
H_{eff}=J\;{\mathbf S}_{1}\cdot {\mathbf
S}_{2}+\gamma_{R}S_{1z}+\gamma_{L}S_{2z}. 
\label{heff} 
\end{equation} 
We use four different approximations to calculate J from the orbital
Hamiltonian given in Eq. (\ref{horb}): Heitler-London (HL), Molecular
Orbital with S states (MO-S), with up to P states (MO-SP), and with up to
D states (MO-SPD). It should be noted that the Zeeman splitting term given
in Eq. (\ref{hzeff}) does not affect the calculation of J in any way and
it is only added to Eq. (\ref{heff}) after J has been determined in order
to obtain the effective 2-spin exchange Hamiltonian controlling the
quantum gates.

\section{Methods and results}

To diagonalize the Hamiltonian given in Eq. (\ref{horb}), we use as our
one particle basis the so called Fock-Darwin energy states centered at the
bottom of the left and right wells ($x=\pm a_{m}$, $y=0$). The Fock-Darwin
QD states, briefly described in Appendix A of this paper for the sake of
completeness and notational clarity, are simply the quantized states of
the 2D orbital motion of the electrons in the presence of an external
parabolic confinement potential and magnetic field. Fock-Darwin states
have a principal quantum number n which describes a shell structure: n=0
is referred to as S states, n=1 as P, n=2 as D, and so on. Each shell is
composed of n+1 levels labeled by the magnetic quantum number m=-n, -n+2,
..., n-2, n. Their energy is given by $E_{nm}^{R/L}=\hbar \Omega
^{R/L}(n+1)+\frac{1}{2}\hbar \omega _{c}^{R/L}m$ where $\Omega
^{R/L}=\sqrt{\omega _{0}^{2}+(\omega _{c}^{R/L}/2)^{2}}$, $\omega
_{c}^{R/L}=eB_{R/L}/mc$ is the cyclotron frequency and $\omega
_{0}=\sqrt{V_{0}/ml_{x}^{2}}\,$ approximates the harmonic confinement
frequency near the bottom of the dot potential. Each Fock-Darwin state has
a characteristic width $l_{0R/L}=\sqrt{\hbar /m\Omega ^{R/L}}$, which
assumes different values depending on whether the state is in the right
well subject to a field $B_{R}$ or in the left well where the field is
$B_{L}$, which are different in the inhomogeneous field case we are
studying.  Notice that the vector potential (\ref{vecpot}) is centered in
between the dots which is located $\pm a_{m}$ to the side of each well.
This means that we have to use basis functions which are gauge transformed
Fock-Darwin states with an additional magnetic phase $\exp{\left(\pm
i\frac{eB_{L/R}a_{m}y}{2\hbar c}\right)}$ (sign $+$ and $-$ refer to the
left and right wells respectively).

In what follows we use this Fock-Darwin single electron basis to form
two-electron states and solve our two-particle problem in two
approximations: Heitler-London which can be solved analytically and
Molecular Orbital which is solved numerically.

\subsection{Heitler-London}

Heitler-London (HL) is the simplest method to calculate the exchange 
coupling J in two coupled dots. It consists of the singlet and
triplet wave functions formed from the
lowest Fock-Darwin states (n=m=0):
\begin{equation}
\Psi _{S/T}(1,2)=\frac{1}{\sqrt{2(1\pm |S|^{2})}}\left( \Psi _{L}(1)\Psi
_{R}(2)\pm \Psi _{R}(1)\Psi _{L}(2)\right) \chi _{s/t},
\label{hlstate}
\end{equation}
\begin{equation}
\Psi _{R/L}({\mathbf r})=\frac{1}{\sqrt{\pi }l_{0R/L}}
\exp{\left(-\frac{(x\mp a_{m})^{2}+y^{2}}{2l_{0R/L}^{2}}\right)}
\exp{\left(\mp i\frac{eB_{R/L}a_{m}y}{2\hbar c}\right)},  
\label{ground}
\end{equation}
where the spin functions are $\chi _{s}=\frac{1}{\sqrt{2}}\left( \left|
\uparrow \downarrow \right\rangle -\left| \downarrow \uparrow \right\rangle
\right) $ and $\chi _{t}$ is one of the three triplet states 
$\frac{1}{\sqrt{2}} \left( \left| \uparrow \downarrow \right\rangle
+\left| \downarrow \uparrow
\right\rangle \right) ,\;\left| \uparrow \uparrow \right\rangle ,\;\left|
\downarrow \downarrow \right\rangle $. The exchange energy is then
calculated as the difference between the expectation values of the
Hamiltonian (\ref{horb}) calculated with the singlet and triplet trial
wave functions: 
\begin{equation}
J=\left\langle T\right| H_{orb}\left| T\right\rangle -\left\langle
S\right|H_{orb}\left| S\right\rangle .
\end{equation}

Using this formula in an inhomogeneous field, we can obtain an exact
expression for J, which is quite complicated when both $\overline{B}$
and $B_{dif}$ are nonzero, and is given in its entirety in Appendix B.
Here we show the expression for J in the special situation when
$\overline{B}=0$,
which catches the main qualitative features of the inhomogeneous case:
\begin{eqnarray}
J(\overline{B}=0,B_{dif})&=&\frac{2S^{2}}{1-S^{4}} \left\{ \frac{\hbar
^{2}a_{m}^{2}}{ml_{0}^{4}}+\frac{1}{4}\frac{e^{2}}{mc^{2}}
B_{dif}^{2}(3l_{0}^{2}-a_{m}^{2})
-\frac{2V_{0}l_{x}l_{y}}{\sqrt{(l_{x}^{2}+l_{0}^{2})(l_{y}^{2}+l_{0}^{2})}}
\right. \label{hl} \\
&&\times\left[
\exp{\left(-\frac{(a-a_{m})^{2}}{l^{2}_{x}+l_{0}^{2}}\right)}
+\exp{\left(-\frac{(a+a_{m})^{2}}{l_{x}^{2}+l_{0}^{2}}\right)}  
-2\exp{\left(-\frac{a^{2}}{l_{x}^{2}+l_{0}^{2}}\right)}
\right] \nonumber \\
&&\left.-\frac{2V_{b}l_{bx}l_{by}}{\sqrt{(l_{bx}^{2}+l_{0}^{2})(l_{by}^{2}+l_{0}^{2})}}
\left[1-\exp{\left(-\frac{a^{2}_{m}}{l_{bx}^{2}+l_{0}^{2}}\right)}\right] 
-\sqrt{\frac{\pi }{2}}\frac{e^{2}}{\varepsilon
l_{0}}\left[1-SI_{0}(\frac{
a_{m}^{2}}{l_{0}^{2}})\right] \right\},  \nonumber
\end{eqnarray}
where $S=\exp{\left(-a_{m}^{2}/l_{0}^{2}\right)}$ is the overlap, $
l_{0}=l_{0R}=l_{0L}$ the Fock-Darwin characteristic width (the same
for the left and right wells in the $\overline{B}=0$ case as $\left|
B_{L}\right| =\left| B_{R}\right| $), and $I_{0}$ the zeroth order
modified
Bessel function. Note that in general $a_{m}\neq a$ since the location of
the minimum in the wells is usually different from the parameter $a=15nm$.
The main difference between Eq. (\ref{hl}) and the homogeneous field 
$B_{dif}=0,\overline{B}\neq 0$
case (see Appendix B) is that the latter contains a number of phase
interference terms
proportional to $\exp \left[ -(e\overline{B}/2\hbar
c)^{2}a_{m}^{2}l_{0}^{2}\right] $, responsible for the faster decay of J
as a function of $\overline{B}$ as will be discussed below. In Fig. 2 we
depict, primarily for the purpose of comparison with the inhomogeneous
field results shown in Figs. 3 and 4, the calculated exchange coupling J
as a function of the external magnetic field B in the homogeneous field
case with $B_{dif}=0$ in all our approximations. 

Two curves calculated for different inhomogeneities using
the HL approximation are shown in Fig. 3. The HL method is
qualitatively reasonable, however, quantitatively it differs significantly
from the more sophisticated molecular orbital calculations described below
in section III B because J depends sensitively on the wave function
overlap, which can be significantly affected by including higher excited
states (as is done in MO calculations, but neglected in the HL theory).
Even though the maximum field gradient obtainable in a laboratory is
around 1 T in a 30nm range, we show values of $B_{dif}$ that go all the
way up to 5 T for completeness. 

\subsection{Molecular orbital calculation}

The basic idea of the MO calculation is to mix in higher excited states in
the set of basis states so as to improve the energetics. To go beyond the
HL approximation, we can assemble more two-particle wave
functions from the Fock-Darwin states, which we call molecular orbitals 
\begin{equation}
\Psi _{i}(1,2)=\frac{1}{\sqrt{2}}\left( \Psi _{i(1)}(1)\Psi _{i(2)}(2)\pm
\Psi _{i(2)}(1)\Psi _{i(1)}(2)\right) \chi _{s/t}\;.
\end{equation}
Electrons are fermions so the singlet has to be symmetric in space
while the triplet is antisymmetric. Here we use this molecular orbital (MO)
basis to solve a generalized eigenvalue problem ${\mathbf H}\cdot {\mathbf \Psi}
=E\;
{\mathbf S} \cdot {\mathbf \Psi }$ where $H_{ij}=\left\langle \Psi _{i}\right|
H\left|
\Psi _{j}\right\rangle $, 
${\mathbf \Psi }=(\Psi_{1}, \Psi_{2},\ldots )$ is
the vector state and the overlap matrix $S_{ij}=<\Psi _{i}|\Psi _{j}>$ is
needed for orthonormalization since the MO basis is in general not
orthonormal. We have done such a
calculation in a constant homogeneous field\cite{xuedong}, using both S and
P Fock-Darwin states. Here we perform calculations with the progressively 
larger basis sets of S, SP, and SPD states in the inhomogeneous field
problem.

In the simplest MO calculation we use only the Fock-Darwin S states to
form the two-electron orbitals. Such a calculation improves upon the HL
method by incorporating the doubly occupied states $\Psi _{R}(1)\Psi_{R}(2)$ 
and $ \Psi _{L}(1)\Psi _{L}(2)$ and we call this the MO-S
calculation. We then have a $3\times 3$ singlet matrix to diagonalize,
and the triplet energy is the same as the HL case (a $1\times 1$ matrix).
Note that we cannot use parity symmetry here to break up the matrices in a
block diagonal form as was done before\cite{xuedong}\ since the field is
inhomogeneous in the x direction. The resulting MO calculation is carried
out numerically. The results of such calculations are shown in Figures 2
(homogeneous case), and 3 (inhomogeneous case) where there is a
noticeable quantitative difference between the HL and MO-S exchange
energies.

To improve the accuracy of our molecular orbital calculations, we also
perform an MO-SP calculation, adding to our basis the $n=1$,
$m=\pm 1$
states on each side so that we have 6 one particle levels in total. The
singlet
MO-SP matrix is $21\times 21$ ($\left( \begin{array}{c} 6\\2
\end{array} \right) +6=21$) and the triplet matrix is $15\times 15$.
We are able to efficiently calculate the Coulomb matrix elements $C_{ij}=
\left\langle \Psi _{i}\right| C\left| \Psi _{j}\right\rangle $ by
reducing the 4-dimensional two-particle integrals to a 1D integral using
the identity 
\begin{equation}
\frac{1}{\left| \mathbf{x}_{1}-\mathbf{x}_{2}\right| }=\frac{1}{r_{12}}=
\frac{1}{\sqrt{\pi }}\int_{0}^{\infty }\frac{ds}{\sqrt{s}}\exp (-sr_{12}^{2})
\end{equation}
and integrating over $\mathbf{x}_{1}$\ and $\mathbf{x}_{2}$ in the
gaussian
basis\cite{shavitt}. 
We then further increase the size of the MO basis by including Fock-Darwin
D states (n=2, m=-2,0,2). We now have 12 single-particle levels, leading
to a $78\times 78$ and a $66\times 66$ two-electron
Hamiltonian matrix in the singlet and triplet basis, respectively.  

Before we analyze our results it is important to determine the range of
validity of our approximations. A strong inhomogeneity
across the region where our basis functions vary appreciably leads to the
necessity of including more and more orbitals to calculate the energy
levels accurately. The extra terms that a
non-vanishing inhomogeneity ($B_{dif}\neq 0$) adds to the one particle
Hamiltonian can be written explicitly as a one particle operator $\Delta
H$. We write the vector potential as ${\mathbf A}={\mathbf A}_{R}+{\mathbf
\delta }_{R}$ where ${\mathbf \delta}_{R}$ represents the deviation from
${\mathbf A}_{R}=-\frac{1}{2}yB_{R} {\mathbf
\widehat{x}}+\frac{1}{2}xB_{R}{\mathbf \widehat{y}}$ which describes the
field in the right well. The one particle Hamiltonian becomes
\begin{equation}
h_{i}=\frac{1}{2m}({\mathbf p}_{i}+\frac{e}{c}{\mathbf A}_{R})^{2}+V+
\underbrace{\frac{e}{mc} {\mathbf \delta}_{R} . ({\mathbf 
p}_{i}+\frac{e}{c}{\mathbf A}_{R})+\frac{1}{2m}(\frac{e}{c})^{2}
\delta _{R}^{2}}_{\Delta H}.
\label{hi}
\end{equation}
$\Delta H$ is the part of the orbital Hamiltonian (\ref{horb}) that is
most sensitive to inhomogeneities in the magnetic field, since it is
proportional to $B_{dif}^{2}$ (the Coulomb matrix elements do not depend
strongly on $B_{dif}$). In an inhomogeneous field, $\Delta H$ strongly
couples the S orbitals with the P and D orbitals, as can be seen from
Table 1, where this orbital mixing is quantitatively represented. Each
individual term in $\Delta H$ contributes approximately the same to the
coupling, so we cannot neglect any of the terms in the calculation.

Note that as $B_{dif}$ increases, the corrections to the MO energy levels
grows very fast, reaching 10\% of the exchange coupling J ($J\cong 0.2meV$
at $\overline{B}=0 T$ and $J\cong 0.1meV$ at $\overline{B}=4 T$) for
$B_{dif}\cong 1 T$ in the second and third columns of Table 1. This
suggests that the HL, MO-S and MO-SP approximations are only reliable up
to $B_{dif}=1 T$ for both the $\overline{B}=0 T$ and $\overline{B}=4 T$
cases. Moreover, a good estimate for the reliability of our best
approximation, MO-SPD can be set as $B_{dif}=4 T$ for the $\overline{B}=0
T$ case and $B_{dif}=2 T$ for the $\overline{B}=4 T$ case. In Fig. 3 we
show the exchange coupling J calculated only up to $B_{dif}=5 T$, which is
an upper limit on the reliability of all our approximations. For
completeness, Fig. 4 displays the same J calculated up to $B_{dif}=25 T$
(In this context, we should emphasize that in realistic experimental
situations $B_{dif}$ is likely to be much less than 1 T).

On the other hand, our approximations are increasingly better at low
inhomogeneity, as can be seen in Table I. Fig. 3 shows that the exchange
coupling J decreases only by 0.1\% in the $\overline{B}=0 T$ case when
$B_{dif}=0.5 T$, which is the maximum field gradient obtainable in an
experiment ($\partial B_{z}/\partial x \cong 300 Gauss/nm$). Also, in the
$\overline{B}=4 T$ case it increases by 0.6\% when the same field gradient
is present. Using these values we can estimate that the error correction
upper bound ($10^{-4}$) corresponds to an inhomogeneity of $30 Gauss/nm$.
In section IV below we will argue that quantum interference in the
inhomogeneous case is responsible for such a small change in J, as
contrasted with the homogeneous situation in Fig. 2.  Moreover, Fig. 3b
shows that the exchange coupling J slightly increases as we switch on the
inhomogeneity in a finite field ($\overline{B}=4 T$). All our
approximations yield an increasing J, which can be understood in a simple
one-particle picture. As we increase $B_{dif}$ the magnetic field on the
right ($B_{R}$) also increases, squeezing the Fock-Darwin state located on
the right quantum dot. However, the left magnetic field ($B_{L}$)
decreases to less than 4 T, which leads to a broader electronic wave
function on the left quantum dot. When $B_{dif}=4 T$, the electronic state
on the left dot reaches its maximum width, and starts to be squeezed for
greater inhomogeneities. This suggests a maximum of the exchange coupling
J located around $B_{dif}=4 T$. This maximum is not evident in our
approximations simply because they lose their accuracy for $B_{dif}$ below
4 T as discussed above, but an increasing trend in J is apparent from Fig.
3b. Moreover, by looking at the $\overline{B}=0 T$ case (Fig. 3a) we can
infer that J is probably not increasing for $B_{dif}<2T$, since, as the MO
approximations become more accurate, J first increase (MO-S and MO-P) and
then decreases (MO-SPD). No maximum in J is expected here since as
$B_{dif}$ increases both the left and right electrons are squeezed.

\section{Discussion}

Comparing the results presented in Figs. 2 and 3 it becomes evident that J
decreases much faster (Fig. 2) in the homogeneous magnetic field, even
becoming negative at around 5T, in both HL and MO approximation schemes.
This behavior can be roughly understood as arising from the approximate
proportionality of J to the wave function overlap $S=\int \Psi_{R}^{*}\Psi
_{L}d{\mathbf r}$ ($\Psi _{R/L}$ are defined in Eq. (\ref{ground}))  
multiplied by a polynomial in $\overline{B}$ and $B_{dif}$\ which explains
features such as the singlet-triplet crossing in Fig. 2 (the overlap
S is plotted in dimensionless units together with our results for J). 
It turns out that $S$ is very sensitive to the magnetic
phase $\exp{\left(\pm i\frac{eB_{L/R}a_{m}y}{2\hbar c}\right)}$ 
which appears in the right and left basis
functions discussed above. In a homogeneous field ($B_{R}=B_{L}=B$) those 
phases add up, leading to destructive interference which gives an
extra exponential decay proportional to
$B^{2}$ in the overlap:
$S_{Hom}=\exp{\left(-(a_{m}/l_{0})^{2}\right)}
\exp{\left(-\left(\frac{1}{2}l_{0}a_{m}\frac{eB}{\hbar
c}\right)^{2}\right)}$. 
On the other hand in the inhomogeneous case with $\overline{B}=0$ ,
$B_{R}=-B_{L}$ the phases cancel out in the S integral leaving just a
single exponential decay,
$S_{Inh}=\exp{\left(-(a_{m}/l_{0})^{2}\right)}$ 
which depends mainly on the
interdot distance. This absence of quantum interference in the
inhomogeneous field case explains why J
hardly changes when there is a magnetic field difference over the two
quantum dots (Fig. 3), which is in fact good news for the operation of the
swap gate in an environment that contains some magnetic impurities for
example (or a field inhomogeneity arising from some other unintended
source).

Note that in the homogeneous case the exchange coupling constant in Fig. 2
decreases extremely strongly at a few Tesla which leads to a
singlet-triplet crossing. This crossing arises from the
competition between the long range Coulomb interaction (which favors the
triplet state) and the kinetic energy (favoring the singlet state). There
is no such singlet-triplet crossing in an inhomogeneous field (except when
$\overline{B}$ is such that $J<0$, where the application of $B_{dif}$
leads to a positive J), because the kinetic energy term (which favors the
singlet state) now dominates as quantum interference terms are
suppressed. In the homogeneous field case on the other hand the potential
energy term always wins at high enough B fields leading to the triplet
state becoming lower in energy than the singlet state.

\section{Conclusion}

In conclusion, we study the singlet-triplet splitting (i.e. the exchange
coupling) J of a horizontally coupled double quantum dot system with two
electrons. In particular, we calculate J when the magnetic field is
inhomogeneous with different field strengths ($\Delta B=2B_{dif}$) in each
dot, and show that realistic values of field inhomogeneity in double
quantum dot structures should have quantitatively small (but
qualitatively interesting) effect on J. We perform an analytical
Heitler-London calculation, and a numerical diagonalization of the orbital
Hamiltonian in a set of increasingly larger two-electron MO basis to
conclude that the cancellation of the magnetic phase difference between
the two electrons on the dots is responsible for the relatively weak
dependence of J on $B_{dif}$ for realistic values of $B_{dif}$. The swap
gate can thus be operated in an inhomogeneous environment as long as
the exchange coupling J is properly chosen and quantum error correction is
performed to account for the inhomogeneity-induced errors arising from the
Zeeman term as discussed earlier\cite{zeeman}. We conclude therefore that
electron spin based QDQC operations should in principle be feasible in an
inhomogeneous magnetic field environment (with the inhomogeneity arising
either from external magnetic fields needed for single qubit manipulation,
or for running the parallel pulse scheme, or from unintended magnetic
impurities or currents invariably present in the system) as long as the
field inhomogeneity is not too large.

This work is supported by ARDA and the US-ONR. RdS acknowledges useful
discussions with Dr. J. Fabian.

\appendix
\section{Fock-Darwin states}

We briefly describe the Fock-Darwin states\cite{fock-darwin}, which are
used as the single
particle basis in our calculations. Consider an electron of mass $m^{*}$
restricted to move in the x-y plane and confined in a parabolic well of
frequency $\omega _{0}$. A magnetic field B is applied in the z direction,
which can be described by the vector potential ${\mathbf A}=-\frac{1}{2}yB
{\mathbf \widehat{x}}+\frac{1}{2}xB{\mathbf \widehat{y}}$. The Hamiltonian then
becomes
\begin{equation}
H=\frac{1}{2m^{*}}\left( {\mathbf p}+\frac{e}{c}
{\mathbf A}\right) ^{2}+\frac{1}{2}m^{*}\omega _{0}^{2}r^{2},
\end{equation}
where $r=\sqrt{x^{2}+y^{2}}$ is the distance between the electron and the origin,
e the electronic charge and c the speed of light. Since this Hamiltonian
commutes with the z component of the angular momentum operator $L_{z}$, its
eigenfunctions can be chosen with well defined angular momentum $m\hbar$.
Apart from the magnetic quantum number m, a radial quantum number n
completely describes the eigenenergies of the system, given by $
E_{nm}=(n+1)\hbar \Omega +\frac{1}{2}m\hbar \omega _{c}$, with $
n=0,1,2,\ldots $ and $m=-n,-n+2,\ldots ,n-2,n$. Here $\omega _{c}=\frac{eB}{
m^{*}c}$ is the electron's cyclotron frequency whereas $\Omega =\sqrt{\omega
_{o}^{2}+(\omega _{c}/2)^{2}}$ is the harmonic oscillator frequency
renormalized by the magnetic field. For small magnetic fields $(\omega
_{c}\ll \omega _{0})$ the quantum number n describes a shell structure, with
$n=0,1,2,\ldots $ being the $S,P,D,\ldots $ shells respectively.  In table 2
below we write the angular momentum and energy eigenfunctions for the S, P
and D shells that are used in calculations throughout this paper. The
negative m wave functions can be easily obtained from the positive ones
using the relation $\phi _{n-m}(r,\theta )=\phi _{nm}^{*}(r,\theta )$. Note   
that a characteristic length $l_{0}=\sqrt{\hbar /m\Omega }$ sets the range
of these wave functions.

\section{Exchange coupling J in the Heitler-London approximation}

Here we write down our analytical results in our calculation of the
exchange coupling J in the Heitler-London approximation. $\overline{B}$
and $B_{dif}$ define the inhomogeneous external magnetic field in the
system (Fig. 1).
As described in section III A, J is calculated by subtracting the average
energies of the Heitler-London states $\Psi _{S/T}$
(Eq. (\ref{hlstate})) leading to 
\begin{equation}
J=\frac{2}{1-\left| S\right| ^{4}}\left\{ \left| S\right| ^{2}\left[ \langle
\Psi _{R}|h|\Psi _{R}\rangle +\langle \Psi _{L}|h|\Psi _{L}\rangle +D\right]
-\left[ 2 Re(S\langle \Psi _{R}|h|\Psi _{L}\rangle) +I\right] \right\} ,
\end{equation}
where h is the single particle Hamiltonian (3), and $S=\langle \Psi
_{L}|\Psi _{R}\rangle $\ is the overlap. We also define the direct (D) and
exchange (I) Coulomb integrals
\begin{equation}
D=\int |\Psi _{L}(1)|^{2}\frac{e^{2}/\varepsilon }
{|{\mathbf x}_{1}-{\mathbf x}_{2}|}|\Psi _{R}(2)|^{2}d{\mathbf x}_{1}
d{\mathbf x}_{2},
\end{equation}
\begin{equation}
I=\int \Psi_{L}^{*}(1)\Psi _{R}^{*}(2)\frac{e^{2}/\varepsilon }
{|{\mathbf x}_{1}-{\mathbf x}_{2}|}\Psi _{L}(2)
\Psi _{R}(1)d{\mathbf x}_{1}d{\mathbf x}_{2},
\end{equation}
where e is the electron charge and $\varepsilon $ the dielectric constant of
the material. We list below all the analytic expressions for those
quantities as a function of the effective mass m, the characteristic lengths
in the right and left wells $l_{0R/L}=\sqrt{\hbar /m\Omega ^{R/L}}$, the
magnetic fields $B_{R/L}=\overline{B}\pm B_{dif}$, and the confining
potential parameters $a$, $l_{x}$, $l_{y}$, $l_{bx}$, $l_{by}$ 
(Eq. (\ref{confpot})). Note
that $a_{m}$ is the minimum location of each well, which is in general
different than the parameter $a$. Throughout the expressions we use the
zeroth order modified Bessel function $I_{0}(x)$, $\gamma =\frac{1}{
l_{0R}^{2}}+\frac{1}{l_{0L}^{2}}$, $\lambda =\frac{1}{l_{0R}^{2}}-\frac{1}{
l_{0L}^{2}}$ , $b=\frac{e\overline{B}}{\hbar c},$ and the overlap $S$, 
which are listed below. 
\begin{equation}
S=\frac{2l_{0R}l_{0L}}{l_{0L}^{2}+l_{0R}^{2}}
\exp{\left(-\frac{2a_{m}^{2}}{l_{0L}^{2}+l_{0R}^{2}}-\frac{b^{2}a_{m}^{2}}{2\gamma}
\right)},
\end{equation}
\begin{equation}
D=\sqrt{\pi }\frac{e^{2}}{\varepsilon }
\frac{1}{\sqrt{l_{0L}^{2}+l_{0R}^{2}}}
\exp{\left(-\frac{2a_{m}^{2}}{l_{0L}^{2}+l_{0R}^{2}}\right)}
I_{0}\left(\frac{2a_{m}^{2}}{l_{0L}^{2}+l_{0R}^{2}}\right), 
\end{equation}
\begin{equation}
I=\sqrt{\pi }\frac{e^{2}}{\varepsilon}
\frac{S}{\sqrt{l_{0L}^{2}+l_{0R}^{2}}}
\exp{\left(-\frac{2a_{m}^{2}}{l_{0L}^{2}+l_{0R}^{2}}\right)}
I_{0}\left( \frac{b^{2}a_{m}^{2}}{2\gamma}\right), 
\end{equation}
\begin{eqnarray}
\langle \Psi _{L}|h|\Psi _{L}\rangle  &=&\frac{\hbar ^{2}}{2ml_{0L}^{2}}+
\frac{1}{8}\frac{e^{2}}{mc^{2}}\left[ \overline{B}^{2}-(B_{R}-B_{L})B_{L}+
\frac{3}{4}\frac{l_{0L}^{2}}{a_{m}^{2}}B_{dif}^{2}\right] l_{0L}^{2} \\
&&-\frac{V_{0}l_{x}l_{y}}{\sqrt{\left( l_{0L}^{2}+l_{x}^{2}\right) 
\left( l_{0L}^{2}+l_{y}^{2}\right) }}
\left[ 
\exp{\left(-\frac{(a_{m}+a)^{2}}{l_{0L}^{2}+l_{x}^{2}}\right)}
+\exp{\left(-\frac{(a_{m}-a)^{2}}{l_{0L}^{2}+l_{x}^{2}}\right)}
\right] \nonumber \\
&&+\frac{V_{b}l_{bx}l_{by}}
{\sqrt{\left( l_{0L}^{2}+l_{bx}^{2}\right) \left(
l_{0L}^{2}+l_{by}^{2}\right) }}  
\exp{\left(-\frac{a_{m}^{2}}{l_{0L}^{2}+l_{bx}^{2}}\right)}.
\nonumber
\end{eqnarray}
For the average of the single particle Hamiltonian h in the right side,
just substitute L for R and vice-versa in the expression above.
\begin{eqnarray}
\langle \Psi _{R}|h|\Psi _{L}\rangle  &=&\frac{1}{2\gamma }S \left\{-\frac{\hbar
^{2}}{m}\frac{1}{l_{0L}^{2}l_{0R}^{2}}\left( \frac{4a_{m}^{2}}{
l_{0L}^{2}+l_{0R}^{2}}-2\right) +\frac{e^{2}a_{m}^{2}}{mc^{2}}\left[\frac{
\gamma }{4}(\overline{B}^{2}+B_{dif}^{2})-\frac{\overline{B}^{2}}{
l_{0L}^{2}+l_{0R}^{2}}-\frac{\lambda }{2}\overline{B}B_{dif}\right] \right. \\
&&+\frac{2}{\gamma }\frac{e^{2}a_{m}^{2}}{mc^{2}}\left[\frac{\overline{B}
^{2}}{l_{0L}^{2}l_{0R}^{2}}+\frac{1}{4}\left( \lambda \overline{B}+\frac{
B_{dif}}{a_{m}^{2}}+\frac{\lambda ^{2}}{\gamma }B_{dif}\right) \left(
\lambda \overline{B}-\gamma B_{dif}\right) \right]   \nonumber \\
&&+\frac{1}{4}\frac{e^{2}\overline{B}^{2}}{mc^{2}}
\left[ 2-\frac{b^{2}a_{m}^{2}}{\gamma}
+\frac{3}{\gamma a_{m}^{2}}\left( \frac{B_{dif}}{\overline{B}}\right)^{2}
+\frac{\lambda}{\gamma}\left( 1+\frac{\lambda }{\gamma }\frac{B_{dif}}
{\overline{B}}\right)
\left( \lambda a_{m}^{2}+\frac{B_{dif}}{\overline{B}}
\left(\frac{\lambda^{2}}{\gamma}a_{m}^{2}+6 \right) 
\right)
\right] 
\nonumber \\
&&-\frac{V_{0}}{\sqrt{\frac{1}{2}
\left( 1+\frac{l_{0L}^{2}}{l_{0R}^{2}}\right) +\frac{l_{0L}^{2}}{l_{y}^{2}}}
\sqrt{\frac{1}{2}\left( 1+\frac{l_{0R}^{2}}{l_{0L}^{2}}\right) +\frac{
l_{0R}^{2}}{l_{x}^{2}}}} 
\exp{\left(-\frac{b^{2}l_{y}^{2}a_{m}^{2}}{2\left(2+\gamma l_{y}^{2}\right)}
-\frac{2l_{x}^{2}a_{m}^{2}}{l_{0L}^{2}l_{0R}^{2}(2+\gamma
l_{x}^{2})}\right)} \nonumber \\
&&\times \left[\exp{\left(-\frac{\left(
a_{m}-a\right) ^{2}/l_{0R}^{2}+
\left( a_{m}+a\right) ^{2}/l_{0L}^{2}}{2+\gamma l_{x}^{2}}\right)}+
\exp{\left(-\frac{\left( a_{m}-a\right) ^{2}/l_{0L}^{2}+\left(
a_{m}+a\right) ^{2}/l_{0R}^{2}}{2+\gamma l_{x}^{2}}\right)}\right]
\nonumber \\
&& \left. +\frac{V_{b}}
{\sqrt{\frac{1}{2}\left( 1+\frac{l_{0L}^{2}}{l_{0R}^{2}}\right) +\frac{l_{0L}^{2}}{
l_{by}^{2}}}\sqrt{\frac{1}{2}\left( 1+\frac{l_{0R}^{2}}{l_{0L}^{2}}\right) +
\frac{l_{0R}^{2}}{l_{bx}^{2}}}}
\exp{\left(-\frac{b^{2}l_{by}^{2}a_{m}^{2}}{2\left(2+\gamma l_{by}^{2}\right)}
-\frac{2\frac{l_{bx}^{2}}{l_{0L}^{2}l_{0R}^{2}}+\gamma }
{2+\gamma l_{bx}^{2}}a_{m}^{2}\right)} \right\}.
\nonumber
\end{eqnarray}

A particularly interesting expression is the homogeneous field exchange
coupling $J(\overline{B},B_{dif}=0)$\cite{guido1}, which we list below (in
addition to Eq. (\ref{hl})) which is the inhomogeneous case
$J(\overline{B}=0,B_{dif})$). Since for this case $B_{R}=B_{L}$, we
simplify our notation by using $l_{0}=l_{0R/L}$ for the characteristic 
length. Also the overlap becomes 
$S(\overline{B},B_{dif}=0)=
\exp{\left(-(a_{m}/l_{0})^{2}\right)}
\exp{\left(-\left( \frac{1}{2}l_{0}ba_{m}\right) ^{2}\right)}$. 
\newpage
\begin{eqnarray}
J(\overline{B},B_{dif}=0)&=&\frac{2S^{2}}{1-S^{4}}\left\{\frac{\hbar
^{2}a_{m}^{2}}{ml_{0}^{4}}-\frac{1}{2}\frac{e^{2}\overline{B}^{2}}{mc^{2}}
a_{m}^{2}\left( 1-\frac{1}{8}b^{2}l_{0}^{4}\right)-\frac{2V_{0}l_{x}l_{y}}{\sqrt{(l_{0}^{2}+l_{x}^{2})(l_{0}^{2}+l_{y}^{2})}}
\right. \\
&&\times \left[ 
\exp{\left(-\frac{(a-a_{m})^{2}}{l_{0}^{2}+l_{x}^{2}}\right)}
+\exp{\left(-\frac{(a+a_{m})^{2}}{l_{0}^{2}+l_{x}^{2}}\right)}
-2\exp{\left(-\frac{a^{2}}{l_{0}^{2}+l_{x}^{2}}+\frac{l_{0}^{2}(bl_{0}a_{m})^{2}}
{4(l_{0}^{2}+l_{y}^{2})}\right)}
\right] \nonumber \\
&&-\frac{2V_{b}l_{bx}l_{by}}{\sqrt{
(l_{0}^{2}+l_{bx}^{2})(l_{0}^{2}+l_{by}^{2})}}
\left[ 
\exp{\left(\frac{l_{0}^{2}(bl_{0}a_{m})^{2}}{4(l_{0}^{2}+l_{by}^{2})}\right)}
-\exp{\left(-\frac{a_{m}^{2}}{l_{0}^{2}+l_{bx}^{2}}\right)}
\right] \nonumber \\
&& \left. -\sqrt{\frac{\pi }{2}}\frac{e^{2}}{\varepsilon l_{0}}e^{(\frac{1}{2}
bl_{0}a_{m})^{2}}\left[ I_{0}\left( \left( \frac{1}{2}bl_{0}a_{m}\right)
^{2}\right) -SI_{0}\left( \left( \frac{a_{m}}{l_{0}}\right) ^{2}\right)
\right] \right\}.  \nonumber
\end{eqnarray}

\newpage

\begin{table}
\begin{tabular}{|c||c|c|c|}
\multicolumn{4}{c}{$\overline{B}=0$ Tesla} \\ \hline\hline
$B_{dif} \ [Tesla]$
& ${\displaystyle  \sum_{n=1,2,3}\sum_{m} \frac{\left|\left\langle
00\right| \Delta H\left|
nm\right\rangle \right|^{2}}{\left| E_{00}-E_{nm}\right|}}$
& ${\displaystyle  \sum_{n=2,3}\sum_{m} \frac{\left|\left\langle
00\right| \Delta H\left|
nm\right\rangle \right|^{2}}{\left| E_{00}-E_{nm}\right|}}$       
& ${\displaystyle  \sum_{n=3}\sum_{m} \frac{\left|\left\langle
00\right| \Delta H\left|
nm\right\rangle \right|^{2}}{\left| E_{00}-E_{nm}\right|}}$ \\ \hline 
0 & 0 meV& 0 meV& 0 meV \\ \hline
1 & 0.041 & 0.037 & 0.002 \\ \hline
2 & 0.163 & 0.143 & 0.011 \\ \hline
4 & 0.744 & 0.591 & 0.053 \\ \hline
6 & 1.95 & 1.37 & 0.156 \\   \hline\hline
\multicolumn{4}{c}{$\overline{B}=4$ Tesla}  \\ \hline\hline
0 & 0 meV& 0 meV& 0 meV \\ \hline
1 & 0.038 & 0.034 & 0.002 \\ \hline
2 & 0.153 & 0.141 & 0.009 \\ \hline
4 & 0.694 & 0.603 & 0.041 \\ \hline
6 & 1.91 & 1.49 & 0.142 \\ \hline
\end{tabular}
\caption{Energy correction of the lowest Fock-Darwin state (S) when P,
D and F levels are included in an inhomogeneous field calculation. 
$\Delta H$ contains all the inhomogeneous field terms that
show up in the one particle Hamiltonian $h_{i}$ (Eq. (\ref{hi})). The
quantum dot
parameters (defined below Eq. (\ref{confpot})) are the same as the ones
used
in Figs. 2, 3 and 4.}
\end{table}

\begin{table}
\begin{tabular}{|ccc||cc|}
& n \hspace{0.3in} m && Eigenstate & \\ \hline \hline
& 0 \hspace{0.3in}  0 && $\frac{1}{\sqrt{\pi
}l_{0}}e^{-\frac{r^{2}}{2l_{o}^{2}}}$ & \\
\hline
& 1 \hspace{0.3in} 1 && $\frac{1}{\sqrt{\pi
}l_{0}^{2}}re^{-\frac{r^{2}}{2l_{o}^{2}} 
}e^{i\theta }$ &  \\  \hline
& 2 \hspace{0.3in} 0 && 
$\frac{1}{\sqrt{\pi}l_{0}}(1-\frac{r^{2}}{l_{o}^{2}})e^{-\frac{r^{2}}{
2l_{o}^{2}}}$ &  \\ \hline
& 2 \hspace{0.3in}  2 && $\frac{1}{\sqrt{2\pi
}l_{0}^{3}}r^{2}e^{-\frac{r^{2}}{2l_{o}^{2}}
}e^{2i\theta }$ &  \\ 
\end{tabular}
\caption{S, P and D Fock-Darwin energy eigenstates for positive
magnetic quantum number m. For negative m, complex conjugate the
corresponding wave function.}
\end{table} 

\begin{figure} 

\vbox to 6.5cm {\vss\hbox to 6cm
 {\hss\
   {\includegraphics{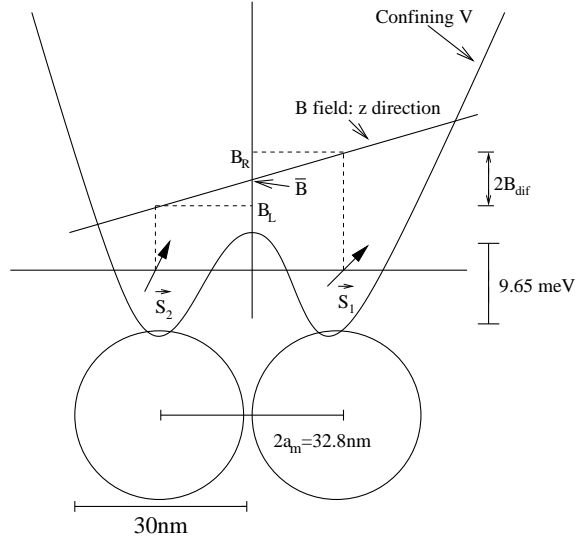}
                }
            \hss}
                }

\caption{Confining potential (defined in Eq. (\ref{confpot}) of the
text) that models
the double dot structure we study. The magnetic field profile is such that
the difference between the fields on the right and left is $2B_{dif}$ and
the average $\overline{B} =(B_{R}+B_{L})/2$ is in the midpoint between
the dots. The quantum dot parameters for all the
calculations in this paper are given in the text following
Eq. (\ref{confpot}).\label{figone}} 
\end{figure}

\begin{figure}

\vbox to 8cm {\vss\hbox to 8cm
 {\hss\
   {\includegraphics{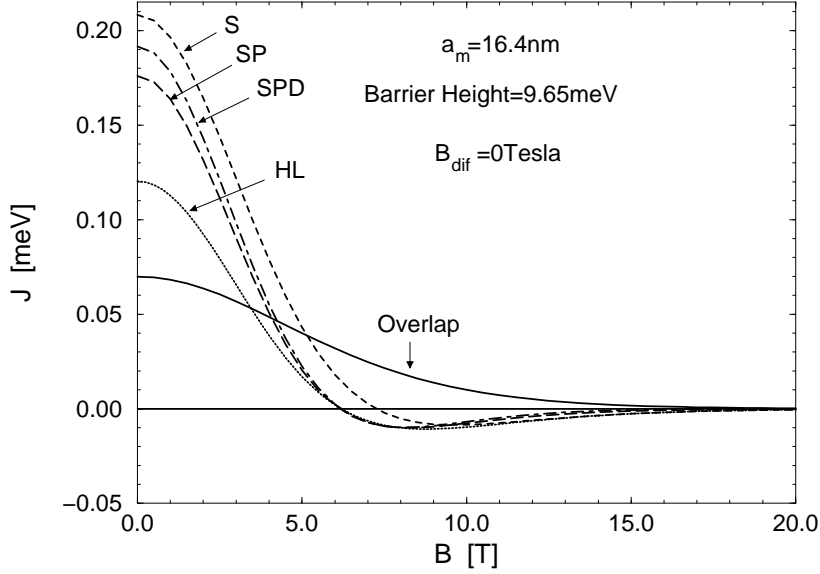}
                }
            \hss}
                }

\caption{Exchange energy J of the coupled dots in a homogeneous
magnetic field $B$. The molecular orbital approximations (S, SP, SPD) are
very close to each other and represent a quantitative improvement over the
simpler Heitler-London (HL). Note that J is plotted in the unit of meV,
while the overlap S is dimensionless. The analytical formula for the HL
theory are given in Appendix B. \label{figtwo}}
\end{figure}

\begin{figure}

\vbox to 8cm {\vss\hbox to 8cm
 {\hss\
   {\includegraphics{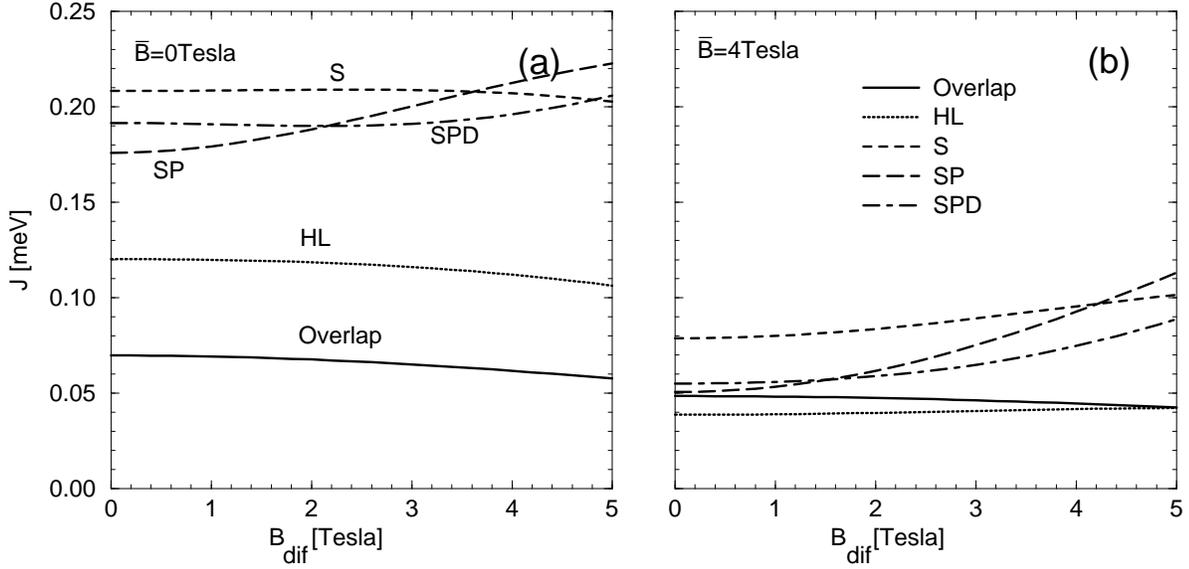}
                }
            \hss}
                }

\caption{Exchange energy J of the double dot structure when an
inhomogeneous magnetic field is present with average $\overline{B}
=(B_{R}+B_{L})/2$ equal to 0 (Fig. 3a) and 4 (Fig. 3b) Tesla respectively.
$B_{dif}=(B_{R}-B_{L})/2$ is half the difference between the fields at the
center of each dot. \label{figthree}} 
\end{figure}

\begin{figure}

\vbox to 8cm {\vss\hbox to 8cm
 {\hss\
   {\includegraphics{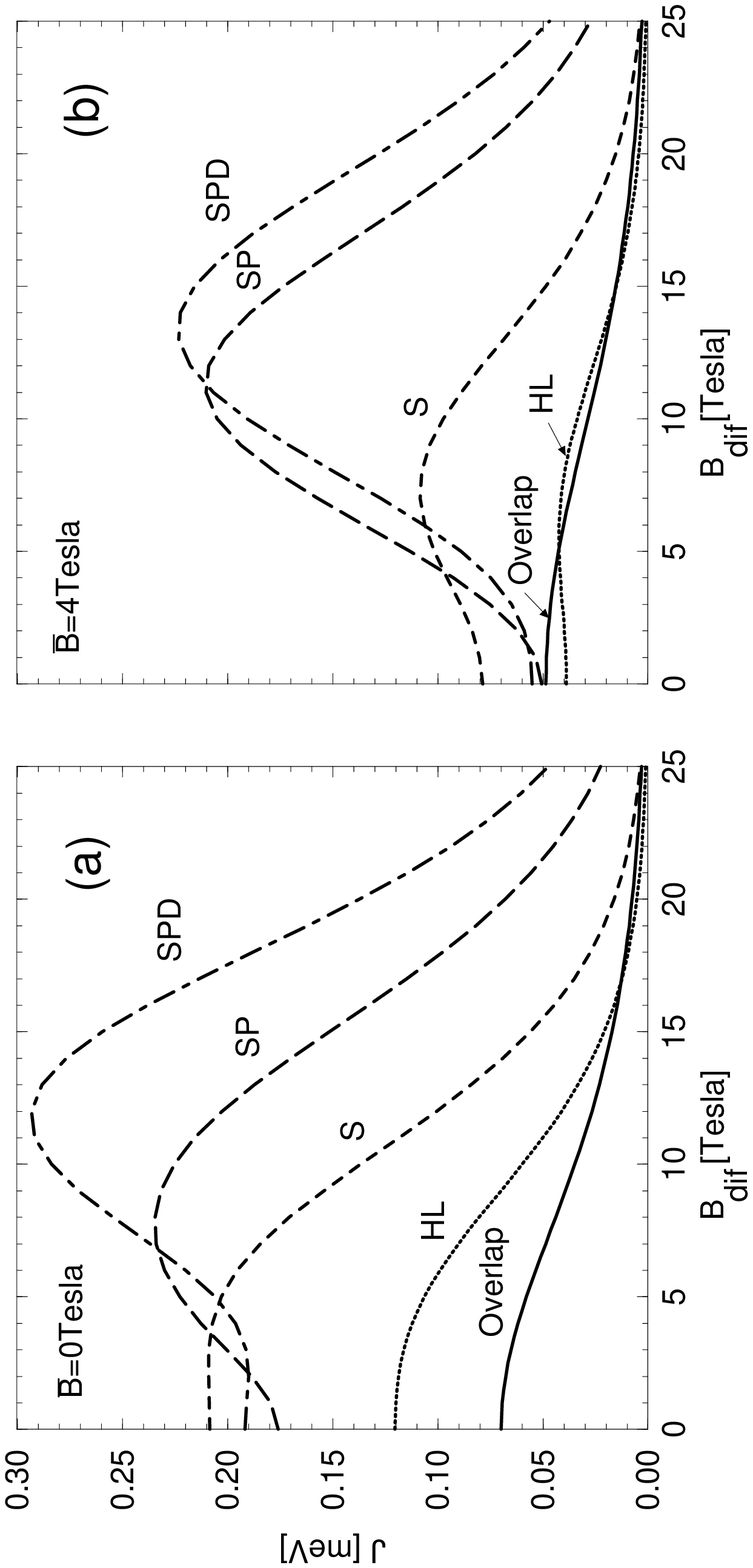}
                }
            \hss}
                }

\caption{Exchange energy J of the double dot when an
inhomogeneous magnetic field is present with average $\overline{B}$ 
equal to 0 (Fig. 4a) and 4 (Fig. 4b) Tesla respectively. For completeness,
we show J as a function of $B_{dif}$ up to 25 T, which is much higher
than what can be obtained in the laboratory. \label{figfour}}
\end{figure}


\begin{thebibliography}{99}
\bibitem{feynman}  R.P. Feynman, Int. J. Theor. Phys. \textbf{21}, 467
(1982); D. Deutsch, Proc. R. Soc. Lond. A \textbf{400}, 97 (1985).

\bibitem{grover}  P.W. Shor, in \textit{Proceedings of the 35th Annual
Symposium on the Foundations of Computer Science} (IEEE Press, Los Alamitos,
CA), p. 124-133 (1994); L.K. Grover, Phys. Rev. Lett. \textbf{79}, 325
(1997).

\bibitem{shor}  P.W. Shor, Phys. Rev. A \textbf{52}, R2493 (1995); A. M.
Steane, Phys. Rev. Lett. \textbf{77}, 793 (1996).

\bibitem{wineland}  C.A. Sacket, D. Kielpinski, B.E. King, C. Langer, V.
Meyer, C.J. Myatt, M.\ Rowe, Q.A. Turchette, W.M. Itano, D.J. Wineland, 
and C. Monroe, Nature \textbf{404}, 256 (2000).

\bibitem{vandersypen} L.M.K. Vandersypen, M. Steffen, G. Breyta, C.S.
Yannoni, R. Cleve, and I.L. Chuang, Phys. Rev Lett. \textbf{85}, 5452
(2000).

\bibitem{braunstein} S.L. Braunstein, C.M. Caves, R. Jozsa, N. Linden,
S. Popescu, and R. Schack, Phys. Rev. Lett. \textbf{83}, 1054 (1999). 

\bibitem{kane}  B.E. Kane, Nature, \textbf{393}, 133 (1998); R. Vrijen,
E. Yablonovitch, K. Wang, H.W. Jiang, A. Balandin, V. Roychowdhury,
T. Mor, and D.P. DiVincenzo, Phys. Rev. A \textbf{62}, 12306 (2000).

\bibitem{loss}  D. Loss and D.P. DiVincenzo, Phys. Rev. A \textbf{57}, 120
(1998).

\bibitem{div1}  D.P. DiVincenzo, Phys. Rev. A \textbf{51}, 1015 (1995).

\bibitem{guido2}  G. Burkard, D. Loss, D.P. DiVincenzo, and J.A. Smolin,
Phys. Rev. B \textbf{60}, 11404 (1999).

\bibitem{zeeman}  X. Hu, R. de Sousa, and S. Das Sarma,
Phys. Rev. Lett. \textbf{86}, 918 (2001).

\bibitem{bacon}  D.P. DiVincenzo, D. Bacon, J. Kempe, G. Burkard, and K.B.
Whaley, Nature, \textbf{408}, 339 (2000); D. Bacon, J. Kempe, D.A. Lidar, 
and K.B. Whaley, Phys. Rev. Lett. \textbf{85}, 1758 (2000); D. Bacon,
K.R. Brown, and K.B. Whaley, Preprint quant-ph/0012018.

\bibitem{guido1}  G. Burkard, D. Loss, and D.P. DiVincenzo, Phys. Rev. B
\textbf{59}, 2070 (1999).

\bibitem{xuedong}  X. Hu and S. Das Sarma, Phys. Rev. A \textbf{61},
62301 (2000).

\bibitem{shavitt}  I. Shavitt, in \textit{Methods in Computational
Physics} Vol 2, Academic Press, N.Y. (1963).

\bibitem{fock-darwin}  V. Fock, Z. Phys. \textbf{47}, 446
(1928); C. Darwin, Proc. Cambridge Philos. Soc. \textbf{27}, 86 (1930).

\end{thebibliography}
\end{document}